\documentclass[twocolumn,showpacs,superscriptaddress,prb,aps,floatfix]{revtex4}
\makeatletter
\def\@dotsep{4.5}
\makeatother

\usepackage{graphicx}
\usepackage{rotating}
\usepackage{amsmath}
\usepackage{color}

\newcommand{\br}{{\bf r}}

\newcommand{\brp}{{\bf r'}}

\renewcommand{\v}[1]{{\bf #1}}
\newcommand{\intint}{\int\!\!\!\int}

\newcommand{\be}{\begin{equation}}
\newcommand{\ee}{\end{equation}}
\newcommand{\bea}{\begin{eqnarray}}
\newcommand{\eea}{\end{eqnarray}}

\begin{document}

\title{A functional of the one-body-reduced density matrix derived from the homogeneous electron gas: Performance for finite systems}

\author{N.\,N.\,Lathiotakis}
\affiliation{Theoretical and Physical Chemistry Institute, NHRF, Vas. Konstantinou 
48, GR-11635, Athens, Greece}
\affiliation{Institut f{\"u}r Theoretische Physik, Freie Universit{\"a}t Berlin, Arnimallee 14, 
D-14195 Berlin, Germany}
\affiliation{European Theoretical Spectroscopy Facility}

\author{N.\,Helbig}
\affiliation{Institut f{\"u}r Theoretische Physik, Freie Universit{\"a}t Berlin, Arnimallee 14, 
D-14195 Berlin, Germany}
\affiliation{European Theoretical Spectroscopy Facility}
\affiliation{Fritz-Haber-Institut der Max-Planck-Gesellschaft, Faradayweg 4-6, 14195 Berlin,
Germany}
\affiliation{Unit\'e de Physico-Chimie et de Physique des Materi\'eaux, Universit\'e Catholique de Louvain, B-1348 Louvain-la-Neuve, Belgium}

\author{A.\,Zacarias}
\affiliation{Institut f{\"u}r Theoretische Physik, Freie Universit{\"a}t Berlin, Arnimallee 14, 
D-14195 Berlin, Germany}
\affiliation{European Theoretical Spectroscopy Facility}

\author{E.\,K.\,U.~Gross}
\affiliation{Institut f{\"u}r Theoretische Physik, Freie Universit{\"a}t Berlin, Arnimallee 14, 
D-14195 Berlin, Germany}
\affiliation{European Theoretical Spectroscopy Facility}

\begin{abstract}
An approximation for the exchange-correlation energy of
reduced-density-matrix-functional theory was recently derived from a
study of the homogeneous electron gas (N.N. Lathiotakis, N. Helbig,
E.K.U.  Gross, Phys. Rev. B {\bf 75}, 195120 (2007)).  In the present work, we
show how this approximation can be extended appropriately to finite systems, where
the Wigner Seitz radius $r_s$, the parameter characterizing the constant density of the
electron gas, needs to be replaced.  We apply the
functional to a variety of molecules at their equilibrium geometry, and also
discuss its performance at the dissociation limit. We demonstrate that, although
originally derived from the uniform gas, the approximation performs
 remarkably well for finite systems. 
\end{abstract}
\maketitle

\section{INTRODUCTION}
Density-functional theory (DFT) is one of the most widely applied tools for
electronic structure calculations. While it successfully describes systems
ranging from atoms and molecules to solids, present day DFT approximations fail
to describe a class of systems generally called {\it strongly correlated}. For
these systems, recent calculations have fueled the hope that
reduced-density-matrix-functional theory (RDMFT) can cure the problem. 
\cite{bb0,gritsenko,sang}
Within RDMFT, the one-body reduced density matrix (1-RDM), $\gamma(\br,\brp)$, is used as
the basic variable in analogy to DFT, where that role is reserved for the
electronic density. The theorem of Gilbert~\cite{gilbert}, which is an
extension of the Hohenberg-Kohn theorem to non-local external potentials,
guarantees that the ground-state expectation value of any observable of a
quantum mechanical system is a unique functional of the ground-state 1-RDM. In
particular, the total energy $E_{\rm tot}$ of a system of $N$ electrons moving
in an external local potential $V(\br)$ can be written in terms of the
ground-state $\gamma$, as 
\begin{multline}
\label{eq:E_of_g}
E_{\rm tot}\left[  \gamma \right] = 
\intint d^3r d^3r^\prime\: \delta(\br - \brp ) 
\left[ -\frac{1}{2} \nabla_{\!\!\br}^2 +V(\br)\right] \gamma(\br, \brp) \\
+\frac{1}{2} \intint d^3r d^3r^\prime \frac{\gamma(\br,\br)\: \gamma(\brp, \brp)}{|\br-\brp|}
+ E_{\rm xc} \left[  \gamma \right] 
\end{multline}
(atomic units are used throughout).
The first term contains the kinetic and external energies and is a simple
functional of $\gamma$. The fact that the kinetic energy can be written
explicitly in terms of the ground-state 1-RDM is a great advantage of RDMFT
compared to DFT. The energy contribution associated with the electron-electron 
interaction can be cast into two terms, the
direct Coulomb energy (or Hartree) term which is again an explicit functional
of  $\gamma$, and the remaining contribution which is called
exchange-correlation (xc) energy. The xc energy is an unknown functional of the
1-RDM and needs to be approximated in practice. Contrary to DFT, however, it does not
contain any kinetic energy contributions but is solely given as the difference
between the full Coulomb interaction and the Hartree energy. Several
approximations for the xc energy have been introduced so
far~\cite{bb0,mueller,sang,UG,csgoe,yasuda,kollmar,kios1,kios2,gritsenko,openshell,pernal,piris,kollmar2,ML2008_1},
the great majority of them being implicit functionals of the
1-RDM. They depend explicitly on the natural orbitals $\varphi_j$ and the
corresponding occupation numbers $n_j$, i.e. the  eigenfunctions and the
eigenvalues of $\gamma$, which are given by
\be
\int d^3r'\gamma(\v r, \v r')\varphi_j(\v r')=n_j\varphi_j(\v r).
\ee
Applications of different RDMFT functionals for the calculation of the
dissociation of molecules~\cite{gritsenko, PLU2007} the ionization
potential~\cite{pirin, LP2006,LP2007, ionpot}, or the fundamental
gap~\cite{LP2007, our_gap}, have been reported. Most approximate RDMFT 
functionals can be written in the form
\begin{multline}
\label{eq:Exc}
E_{\rm xc} \left[  \gamma \right] = E_{\rm xc} \left[ \{ n_j \}, \{ \varphi_j \} \right]=\\
-\frac{1}{2} \sum_{j,l=1}^{\infty}\! f(n_j,n_l)\intint d^3r d^3r^\prime
\frac{\varphi_j^*(\br)\: \varphi_l^*(\brp)\: \varphi_l(\br)\:
\varphi_j(\brp)}{|\br-\brp|} \,,
\end{multline}
with some function $f(n_j,n_l)$ which distinguishes these 
approximations from the Hartree-Fock approximation.

The first approximation of the kind in Eq. (\ref{eq:Exc}) was introduced by
M\"uller~\cite{mueller} using the function $f(n_j,n_l) = \sqrt {n_j n_l}$.
Buijse and Baerends rederived the same approximation from
modeling exchange and correlation holes.~\cite{bb0} The M\"uller functional overcorrelates in all systems it was 
applied to.~\cite{staroverov,herbert,ML_2008_2,ciospernal,csanyi,our_heg}
Interestingly,  for diatomic molecules, it reproduces the dissociation into two
independent atomic fragments correctly. 
Goedecker and Umrigar~\cite{UG} (GU) considered a functional with the same
function $f$ but excluding explicitly the self-interaction terms, $j=l$, in
Eq.~(\ref{eq:Exc}) and in the direct Coulomb term in Eq.~(\ref{eq:E_of_g}).
They used their functional in a minimization
procedure and determined the natural orbitals and the corresponding occupation
numbers. In this way, they found correlation energies, for small atomic and
molecular systems, that are in very good agreement with the exact ones. However, it
was found later that the GU functional fails to describe the dissociation of
diatomic molecules correctly~\cite{staroverov,herbert}.  
In an attempt to correct the overcorrelation of the M\"uller functional, while keeping
the good description in the dissociation limit, Gritsenko et al.~\cite{gritsenko}
proposed a hierarchy of three levels of repulsive corrections. All three
corrections distinguish between weakly and strongly occupied orbitals, the former being
orbitals with occupation numbers close to zero, the latter having occupations close to one.
The resulting functionals are called BBC1,  BBC2, and BBC3.
For the first two approximations the function $f$ is given by
\begin{equation}
f^{\rm BBC1}(n_j,n_l) = 
\left\{  
\begin{array}{rl}
-\sqrt{n_j\: n_l}\,, & j\neq l, \mbox{and\ } \\
 & j,l\mbox{\ weakly occupied,} \\
 \sqrt{n_j\: n_l}\,, & \mbox{\ otherwise,} 
\end{array}\right.
\end{equation}
\begin{equation}
f^{\rm BBC2}(n_j,n_l) = 
\left\{  \begin{array}{rl}
-\sqrt{n_j\: n_l}\,, & j\neq l, \mbox{and\ } \\
 & j,l\mbox{\ weakly occupied,} \\
       n_j\: n_l\,, & j\neq l, \mbox{and\ } \\
 & j,l\mbox{\ strongly occupied,} \\
 \sqrt{n_j\: n_l}\,, & \mbox{\ otherwise.} 
\end{array}\right.
\end{equation}
In the BBC3 functional, the anti-bonding orbital is treated as strongly occupied. 
Additionally, the self-interaction terms are removed  as in the GU functional, except for the
pair of bonding and anti-bonding orbitals. Gritsenko et al used the BBC
functionals to calculate the dissociation curves of diatomic molecules.
 They concluded that the BBC3 functional is very accurate in the description of
these systems both at the equilibrium geometry and the dissociation limit.
Two other functionals, derived from a cumulant expansion of the
second order density matrix, with a final form that is very similar to the BBC
functionals were introduced by Piris.\cite{piris} We refer to these functionals 
as Piris natural orbital functionals. The first approximation, PNOF0, is
identical to the BBC1 functional apart from the self-interaction terms which are
removed in the same way as in the GU approximation. In the second functional,
PNOF, an additional term is included to avoid occupation numbers which are
identical to zero or one. PNOF0 and PNOF coincide for two electron
systems. The BBC as well as PNOF and PNOF0 
functionals were evaluated recently for a large set of molecular systems and
were proven to be quite accurate in reproducing the correlation and the 
atomization energies of these systems.\cite{ML_2008_2}


For the application of 1-RDM functionals to periodic systems, the homogeneous
electron gas (HEG) is an important prototype system.  Also, as far as size
is concerned, the HEG and small atomic and molecular systems are two
opposite extremes. For the HEG, the GU and M\"uller functionals are identical
since the self-interaction terms vanish. Similarly, all terms that include a
special treatment of single orbitals vanish.  As a result, the BBC3 functional
coincides with BBC2, and PNOF0 with BBC1 in this special case.
Cioslowski and Pernal~\cite{ciospernal} as well
as Cs\'anyi and Arias~\cite{csanyi} applied the M\"uller functional to the HEG.
As for finite systems, the correlation energy is overestimated, actually by a
factor of about two in the high density regime. In the low density region, the
M\"uller functional fails completely to reproduce the limit of zero
correlation.\cite{ciospernal,csanyi,our_heg} Cs\'anyi and Arias also introduced
a different functional starting from a tensor expansion of the second-order
matrix. Unfortunately, it fails to describe the electronic correlation of the
HEG in both the dense and dilute limits. A more successful functional for the
HEG is the one proposed by Cs\'anyi, Goedecker and Arias (CGA)~\cite{csgoe}, which
reproduces relatively accurately the correlation energy for the dense HEG. 
Recently, we applied the BBC1 and BBC2 functionals to the HEG~\cite{our_heg}. We
showed that these functionals offer a better description of the correlation of
the HEG over the whole range of densities than any other of the discussed
functionals. Both the BBC1 and the BBC2 functionals overcorrelate slightly for
high densities and undercorrelate for low densities. The crossover is at around
$r_s=0.5$ where these functionals perform best. Additionally, they produce a
finite discontinuity in the momentum distribution at the Fermi energy,
resembling a feature of the exact theory. Unfortunately, the size and the
dependence on the density of this discontinuity are not in agreement with the
exact result. 

\section{THE BBC++ FUNCTIONAL}
In an attempt to improve over the BBC functionals for the HEG, we introduced a
modification to the BBC1 functional~\cite{our_heg}. It consists of the
introduction of a parameter into the function $f$ fitted for each value
of $r_s$ to reproduce the exact correlation energy of the HEG. Two choices for
this parameter were made. We concluded that a reasonable way to generalize BBC1
is to introduce a function $s(r_s)$ multiplying the function $f$ 
when both orbitals are weakly occupied. The corresponding function
$f$ then reads
\begin{equation}
f(n_j, n_l) = 
\left\{  
\begin{array}{rl}
-s(r_s)\: \sqrt{n_j\: n_l}\,, & j\neq l, \mbox{and\ } \\
 & j,l\mbox{\ weakly occupied,} \\
 \sqrt{n_j\: n_l}\,, & \mbox{\ otherwise.} 
\end{array}\right.
\label{eq:ourfunct}
\end{equation}
We call this functional $s$-functional.
Note that, for the HEG, if one assumes plane-wave natural orbitals,
no special care is necessary for the $j=l$ terms since these terms vanish. The
$s$-functional
reproduces, by construction, the exact correlation energy of the HEG. It also
improves the calculated momentum distribution compared to BBC1 and BBC2.
However, the momentum distribution still deviates from the exact one~\cite{our_heg}.

Being derived from the study of the HEG, the $s$-functional is directly
applicable to metallic systems, where a value of $r_s$ can be associated. The
application to other systems is not straightforward since one has to overcome
the $r_s$ dependence. One possibility, in the spirit of the local density
approximation of DFT, is to express $s$ as a function of the local electronic
density. Hence, $s$ becomes a function of the space coordinate $\br$ and
therefore  should be properly included in the integrals in Eq.~(\ref{eq:Exc}). A
reasonable choice that preserves the symmetry of the integrations over $\br$ and
$\brp$ is to multiply the integrand in Eq.~(\ref{eq:Exc}) by $\sqrt{s(n(\br))\:
s(n(\brp))}$ or other possible averages and keep the BBC1 form for the function $f$. 
Work investigating the performance of the resulting functional is in progress.

\begin{figure}
\includegraphics[width=0.9\columnwidth,clip]{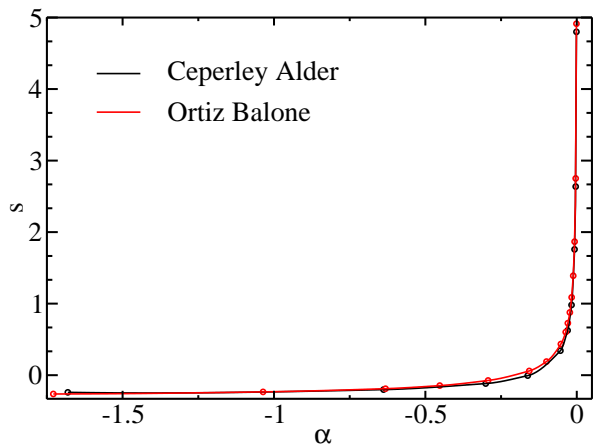}
\caption{\label{fig:s_a} Dependence of the fitting parameter $s$ of the $s$-functional~\cite{our_heg} 
on the ratio $\alpha$ of the correlation over the kinetic energy for the HEG. Two different functions
$s$ are plotted for two sets of diffusion Monte-Carlo data which $s$ was fitted to 
reproduce: From Ceperley and Alder~\cite{exactmom1} and Ortiz and Balone~\cite{exactmom2}.}
\end{figure}

In the present article, we propose an alternative way to circumvent the $r_s$
dependence. We refer to the resulting functional as the BBC++ functional. 
In this approximation, the xc terms
retain the simple form of  the exchange integrals over Gaussian or Slater type
orbitals. The idea is to establish the dependence of $s$ on a quantity $\alpha$
which, contrary to $r_s$, is meaningful for all systems, finite and periodic,
metallic and insulating. We choose the ratio of the correlation energy over the
Hartree-Fock kinetic energy for this quantity $\alpha$. For the HEG, $\alpha$
depends on the density parameter $r_s$ and, since the function $\alpha(r_s)$ is
strictly monotonic and therefore invertible, the dependence $s(\alpha)$ can be
established. The resulting function $s(\alpha)$ for the HEG is shown in 
Fig.~\ref{fig:s_a}. The function $s(r_s)$, which is necessary in the calculation
of $s(\alpha)$, is given in Ref.~\onlinecite{our_heg}. 
The difficulty  for the application of the BBC++ functional lies in the fact
that the correlation energy, and consequently $\alpha$, are only known at the
solution point, i.e. when the optimal $\gamma$ is known.  Thus, $s(\alpha)$ has
to be determined self-consistently during the minimization procedure. Starting
from a trial value for $s$ one minimizes the energy with respect to $\gamma$,
calculates $\alpha$ and feeds the corresponding value for $s$ back into the
functional. As the BBC++ functional coincides with the $s$-functional for the
HEG, the self-consistent determination of $s(\alpha)$ has to yield the correct
$s(r_s)$ in this case. Using the implementation for the HEG presented in 
Ref.~\onlinecite{our_heg}, we verified that this is indeed the case. We also
found that the self-consistent determination of $s$ converges for all the finite
systems we studied.

\begin{table*}[t]
\setlength{\tabcolsep}{0.3truecm}
\begin{tabular}{c|cccccc|c}
\hline\hline
Functional    &     He     &   Be     &     H$_2$    &    LiH    &     Ne    &     HF  & $\bar{\Delta}$\\
\hline
Hartree-Fock  & 2.8615     &  14.5729    &  1.1330   &  7.9868   & 128.5320  & 100.0589 & 0.119 \\
M\"uller      & 2.9143     &  14.7471    &  1.1905   &  8.1167   & 128.9168  & 100.5185 & 0.090 \\
GU            & 2.8980     &  14.6456    &  1.1660   &  8.0478   & 128.8260  & 100.3692 & 0.019 \\
BBC1          & 2.9042     &  14.6692    &  1.1746   &  8.0642   & 128.8471  & 100.4105 & 0.035 \\
BBC3          & 2.9005     &  14.6490    &  1.1705   &  8.0444   & 128.8014  & 100.3373 & 0.011 \\
PNOF          & 2.8925     &  14.6234    &  1.1593   &  8.0310   & 128.7876  & 100.3166 & 0.013 \\
PNOF0         & 2.8925     &  14.6236    &  1.1593   &  8.0309   & 128.7840  & 100.3133 & 0.014 \\
BBC++         & 2.9035     &  14.6574    &  1.1768   &  8.0572   & 128.7862  & 100.3526 & 0.018 \\
CCSD(T)       & 2.9025     &  14.6186    &  1.1724   &  8.0227   & 128.8054  & 100.3401 & 0.000 \\
\hline
$s$           & 1.202      &  2.108     &  0.573     &  1.674    & 3.158     & 2.823    &      \\
$\alpha$      & -0.0147    & -0.00580   & -0.0388    & -0.00882  & -0.00198  & -0.00344 &      \\
\hline\hline
\end{tabular}
\caption{\label{tab:1} 
Absolute total energies (atomic units) of atoms and diatomic molecules calculated with different 1-RDM 
functionals and the average absolute deviation $\bar{\Delta}$ from the reference energies obtained
with the CCSD(T)~\cite{ccsdt} method. For the latter, we used the Gaussian 03 program~\cite{gaussian03}
for the same basis sets. In the last two rows, the values of the optimal values 
of $s$ for the BBC++ functional and the corresponding values of $\alpha$ are also given.
}
\end{table*}

\section{RESULTS FOR FINITE SYSTEMS}

For the application of 1-RDM functionals to finite systems, we used the
implementation introduced in Ref.~\onlinecite{openshell} which relies on the
GAMESS computer program~\cite{gamess} for the calculation of the one- and two-
electron integrals. In the present work, we employed the cc-pVTZ basis
set~\cite{ccpvtz} for all systems apart from the He atom  for which the cc-pVQZ
set~\cite{ccpvqz} was used. Depending on the system, these basis sets contain
30 to 50 basis functions. We always made full use of the size of the basis set,
optimizing as many natural orbitals as there were basis functions available. 
For the BBC3 functional, we used the form that respects the possible 
degeneracies of the bonding and anti-bonding orbitals.\cite{ML_2008_2} The
total energies resulting from this full minimization with respect to the natural
orbitals and occupation numbers for several atoms and diatomic molecules are
given in Table~\ref{tab:1}. We compare all our results to total energies
obtained from a coupled-cluster-singles-doubles-triples (CCSD(T))~\cite{ccsdt} calculation
using the Gaussian 03 computer program~\cite{gaussian03} with the same basis sets as used
in the RDMFT calculation. For the systems considered here, the BBC++ functional
yields slightly better total energies than GU but does not reach the accuracy of
the BBC3 and Piris functionals. 
However, it performs significantly better than the BBC1
approximation that it was derived from, except for the H$_2$ molecule. The
repulsive correction to the BBC1 functional increases with increasing number of
electrons in the system. Overall, for small finite systems, the BBC++ functional performs remarkably well
considering that it was originally tuned to be exact for the HEG.

\begin{figure}[t]
\includegraphics[width=0.9\columnwidth,clip]{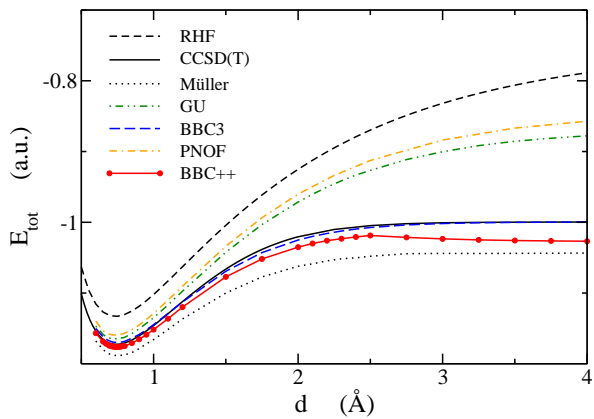}
\caption{\label{fig:stretchedH2} Dissociation curve of the H$_2$ molecule with different 
functionals of the 1-RDM.}
\end{figure}

In Fig.~\ref{fig:stretchedH2}, we plot the dissociation curve for the H$_2$
molecule. As already mentioned, among all the systems we
studied, H$_2$ is the only case where the BBC++ functional does not improve over
BBC1. Nevertheless, BBC++
reproduces the dissociation curve surprisingly well giving a
qualitatively correct optimal 1-RDM with four occupations equal to 0.5 (per
spin) in the limit of large distance. The slight decrease in the total energy
at around 2.5\AA~separation, which can be seen in Fig.~\ref{fig:stretchedH2}, is
a pathology of the BBC++ functional. It originates from the dramatic increase in
static correlation leading to a large negative value for $\alpha$. While
$\alpha$ has a value of -0.0388 at the equilibrium distance, it becomes -0.335
at a distance of 4 \AA. As we can see from Fig.~\ref{fig:s_a}, this results in a
negative value for $s$ driving the BBC++ towards the M\"uller functional.
Consequently, the functional overcorrelates and leads to a decrease in the total
energy with increasing distance for the H$_2$ molecule. It is worth mentioning
that the dissociation of the H$_2$ molecule is a rather difficult case for DFT
functionals~\cite{gruning,baerends}. 

Finally, a second pathology of the BBC++ functional is its obvious size inconsistency: 
Consider a system consisting of two independent sub-systems, for example
two finite systems at a large distance.  If the two sub-systems are
identical, the $\alpha$ of the total system is equal to the values of each
sub-system. In the extreme case, however, where one of the sub-systems is much
larger than the other, the common value for $\alpha$ is completely determined 
by the larger sub-system. That is, since the two systems are independent, the
functional, when applied to the composite system, gives a different result for
the smaller sub-system than when applied to the sub-systems independently.

\section{CONCLUSION}
We presented a RDMFT functional, which we call BBC++, based on an idea to 
circumvent the dependence on the density parameter $r_s$ of functionals derived 
from the homogeneous electron gas. 
This idea is applied to the $s$-functional introduced in 
Ref.~\onlinecite{our_heg}.  
We apply BBC++ in the calculation of correlation energies of small atomic
and molecular systems and show that its performance is satisfactory. We also discuss
pathologies of this functional, with the most important being its size inconsistency.

Despite its pathologies, the BBC++ functional represents an important step in
the development of 1-RDM functionals. It is a successful attempt to apply
approximations originally developed for the HEG to finite systems. In other
words, within RDMFT it is possible to develop functionals that perform equally
well for extended systems, like the HEG, as well as small atomic and molecular
systems. The present work serves as an initiative for the development of better
approximations based on the HEG and, furthermore, their application to finite
systems in the future.

%

\begin{acknowledgments}
This work was supported in part by the  Deutsche Forschungsgemeinschaft within the program SPP 1145,
by the EXCITING Research and Training Network, and by the 
EU's Sixth Framework Program through the Nanoquanta Network of Excellence (NMP4-CT-2004-500198).
\end{acknowledgments}

\end{document}